# Deep Reinforcement Learning of Transition States

Jun Zhang[1,#], Yao-Kun Lei[2,#], Zhen Zhang[3], Xu Han[2], Maodong Li[1], Lijiang Yang[2], Yi Isaac Yang[1,*] and Yi Qin Gao[1,2,4,5,*]

[1] *Institute of Systems and Physical Biology, Shenzhen Bay Laboratory, 518055 Shenzhen, China*
[2] *Beijing National Laboratory for Molecular Sciences, College of Chemistry and Molecular Engineering, Peking University, 100871 Beijing, China.*
[3] *Department of Physics, Tangshan Normal University, 063000 Tangshan, China.*
[4] *Beijing Advanced Innovation Center for Genomics, Peking University, 100871 Beijing, China.*
[5] *Biomedical Pioneering Innovation Center, Peking University, 100871 Beijing, China.*

[#] These authors contributed equally to this work.
[*] Correspondence should be sent to yangyi@szbl.ac.cn (Y.I.Y) or gaoyq@pku.edu.cn (Y.Q.G).

**Abstract**

Combining reinforcement learning (RL) and molecular dynamics (MD) simulations, we propose a machine-learning approach (RL$^‡$) to automatically unravel chemical reaction mechanisms. In RL$^‡$, locating the transition state of a chemical reaction is formulated as a game, where a virtual player is trained to shoot simulation trajectories connecting the reactant and product. The player utilizes two functions, one for value estimation and the other for policy making, to iteratively improve the chance of winning this game. We can directly interpret the reaction mechanism according to the value function. Meanwhile, the policy function enables efficient sampling of the transition paths, which can be further used to analyze the reaction dynamics and kinetics. Through multiple experiments, we show that RL$^‡$ can be trained *tabula rasa* hence allows us to reveal chemical reaction mechanisms with minimal subjective biases.

# I. Introduction

Although chemical reactions are among the most familiar events to chemists, many basic questions remain to be answered, among which the reaction mechanisms are often of particular interest and importance. Being able to explain and predict reaction mechanisms will add insights to the relevant theories, and is essential for manipulating reactions and designing catalysts. One widely applied approach in the study of reaction mechanisms is to introduce reaction coordinate(s) (RC)[1-4] and transition states (TS),[1] which give rise to a simplified description of reduced dimensionality and serve as cornerstones in many successful theories in chemical physics such as the well-known transition state theory,[1, 5] Kramers' theory,[4, 6] electron-transfer theory,[7] etc. However, since real-world chemical processes usually take place in a high-dimensional space which include redundant coordinates apart from those delimiting reactants and products, extracting the relevant factors that characterize the reaction can be very difficult.[4, 8-10] As a consequence, in practice investigation of reaction mechanisms usually starts from some empirical guesses of a plausible mechanism or reaction coordinate according to "chemical intuitions" or expertise hence inevitably involves subjective biases or artifacts.

With the advent of supercomputers, atomistic and multi-scale molecular simulations, e.g., quantum mechanics/ molecular mechanics (QM/MM) MD, have been widely adopted to study chemical reactions.[11] In spite of many successes, MD is limited by its inability to describe long-time scale dynamical processes when a high (free) energy barrier is encountered as in chemical reactions. Since chemical reactions are characterized by a long waiting time inside the local minima (which define the reactant and product) and an ultrashort transition time over the in-between barrier, MD often ends in over-redundant sampling of conformational basins at the cost of insufficient sampling of the barrier region. Therefore, a method is desired to sample efficiently over the transition states, meanwhile assuming least *a priori* assumptions about the reaction coordinates. Particular examples towards such goal include but are not limited to transition path sampling (TPS)[12] and enhanced sampling of reactive trajectories (ESoRT).[13-14] However, since there lacks of a built-in and unsupervised way to learn from the data (e.g., simulation trajectories), the sampling efficiency of these existing methods cannot be improved or bootstrapped automatically.

Deep learning,[15-16] on the other hand, may offer new solutions to the above-mentioned problems due to its ability to extract and generalize features and patterns from high-dimensional data. Particularly, performance achieved by some deep learning models in certain specific tasks is unequivocally superhuman such as the Go-playing computer program AlphaGo,[17] showing that, instead of crafting ever more sophisticated methods relying on empirical expertise, we may train themselves to learn the rules from the data. Equipped with path sampling techniques, we can generate large amount of informative data (e.g., simulation trajectories) and receive feedbacks, so we may cross-fertilize statistical mechanics with deep RL[18] in order to investigate reaction mechanisms. Specifically, as transition states can be regarded as certain intermediates between the reactant and product, they can be defined as an ensemble of configurations which have an equal probability of reaching reactant or product regions.[2, 19-22] We thus formulate a novel game according to this definition and solve it with RL. This game along with the proposed approach to solving it is called Reinforcement Learning of Transition States (RL‡), and RL‡ is able to discover chemical reaction mechanisms presuming limited *a priori* expertise and with minimal subjective biases.

## II. Methods

### 1. RL[‡]: find transition states through a "shooting game"

In RL[‡], locating the transition state of a chemical reaction is formulated as a "shooting game" as follows: We randomly choose a coordinate from the phase space according to a certain distribution (e.g., Maxwell-Boltzmann distribution), then shoot two paired trajectories with opposite momenta from this initial coordinate under micro-canonical ensemble.[3, 23] We end any of the trajectories once it reaches the product (or reactant) region or it exceeds the maximal allowed simulation time (Algorithm S1). Now consider the outcomes of the shooting move: if one of the paired trajectories ends in the reactant region while the other in the product, we win the game; otherwise we lose it. Intuitively, if we could pinpoint the transition state (or the dividing surface) and shoot from configurations therein, we would achieve the highest chance of victory. The "shooting" game played by RL[‡] obeys clear rules and yields binary outcomes. All the knowledge we need to define the game is the Hamiltonian of the system, and an unambiguous definition of the reactant and product. However, this game is very challenging in that the TS occupies only a tiny fraction of the entire configuration space so the positive feedbacks are sparse. Besides, the shooting moves are intrinsically stochastic. Therefore, inspired by reinforced imitation learning, we implement Expert Iteration (EXIT) to conquer it.[17, 24] EXIT employs a pair of Expert and Apprentice to iteratively improve the chance of winning the game (Algorithm S2).

### 2. Expert: Enhanced Sampling of Reactive Trajectories (ESoRT)

The role of the Expert is to perform sampling and thereby to find promising initial configurations which may lead us to victory. According to the Boltzmann distribution, the probability for the Expert to shoot from a certain configuration **R** is proportional to,

$$p_{\text{shoot}} \propto \exp\left(-\beta\left(U(\mathbf{R}) + V(\mathbf{R})\right)\right) \quad (1)$$

with the momentum randomly generated according to Maxwell distribution given the inverse temperature $\beta$ and the potential energy function $U(\mathbf{R})$. Note that a bias term $V(\mathbf{R})$ is included as a "policy function" to manipulate the shooting probability ($p_{\text{shoot}}$) in the spirit of ESoRT[13-14] and importance sampling.

As in any other reinforcement learning tasks, there is a trade-off between exploration and exploitation when choosing the policy function. Exploration means that the sampling should not only focus on the configurations of high predicted values, but should also explore more towards less-visited regions of which the values may be underestimated. In other words, exploration is necessary to avoid being trapped in sub-optimal solutions. As for exploitation, if we know the exact reaction coordinate and transition state, we can replace $V(\mathbf{R})$ by a harmonic potential as in umbrella sampling.[25] In this scenario, RL[‡] is analogues to TPS from the top of the barrier.[3, 23] However, in practice, RL[‡] is able to automatically extract reaction coordinate and learn a bias potential that achieves the same goal when we presume neither the reaction coordinate(s) nor the functional form of the bias potential. In order to play the game, we can perform MD or Monte Carlo sampling on the modified potential energy surface (PES) according to Eq. (1), then shoot trajectories from the randomly sampled configurations. Although the initial configurations are sampled according to a modified distribution, the trajectories are shot under the original Hamiltonian. In this way, no matter how we change the policy function (i.e., the bias potential), the transition paths and reaction mechanisms will remain authentic.

## 3. Apprentice: Actor-Critic Learning

The role of the Apprentice is to summarize and generalize the promising initial coordinates discovered by the Expert across the whole configuration space, and to bias the future sampling towards the potentially promising configurations (i.e., the transition states). The Apprentice consists of two functions, one for value estimation (hence called value function) and one for policy making (hence called policy function), in the spirit of the Actor-Critic algorithm.[26-27]

In RL[‡], the value function estimates the winning chance (i.e., the probability of a successful shooting) from an initial configuration $\mathbf{R}$, which can also be interpreted as the "transition path probability", $p(\text{TP}|\mathbf{R})$. Configurations maximizing $p(\text{TP}|\mathbf{R})$ constitute the separatrix (or TS) of the reaction.[3] Therefore, we can directly interpret the reaction mechanism via a well-trained value function. We adopt an artificial neural network (ANN) with a sigmoid output layer to approximate the value function $p_w(\text{TP}|\mathbf{R})$, where $w$ denotes optimizable parameters. However, we cannot treat the training of $p_w$ as an ordinary regression problem due to two reasons: i) The shooting outcomes are usually highly imbalanced, that is, the failed shootings overwhelm the successful ones; ii) The fate of trajectories started from a same configuration $\mathbf{R}$ is non-deterministic due to randomized initial momenta. Instead, we can optimize $p_w(\text{TP}|\mathbf{R})$ by minimizing the following contrastive loss (see SI for more details),

$$\mathcal{L}(w; \mathbf{R}_i) = -\log \frac{p_w(\mathbf{R}_i)}{p_w(\mathbf{R}_i) + \sum_{j=1,N} p_w(\mathbf{R}_j)} \quad (2)$$

where $\mathbf{R}_i$ is an initial configuration corresponding to a successful shooting while $\mathbf{R}_j$ corresponds to failed ones, and $p_w(\mathbf{R})$ is short for $p_w(\text{TP}|\mathbf{R})$. Equation (2) effectively balances the positive and negative feedbacks, and allows us to optimize $p_w$ by shooting merely one or a few trajectories from a given configuration. In practice, for one successful shooting, we sample $N$ failed shootings to calculate $\mathcal{L}(w)$. We also implemented experience replay[28] to exploit more of the sparse positive feedbacks. More details and further regularization of the value network can be found in SI.

The policy function $V(\mathbf{R})$ in Eq. (1) is used to manipulate the Expert's sampling distribution. There are many optional functional forms for the policy function (see SI for more details), which will be referred as $V_\theta(\mathbf{R})$ hereafter with $\theta$ standing for optimizable parameters. The parametrized $V_\theta(\mathbf{R})$ can be updated through the Policy Gradient Optimization (PGO) algorithm[29] if one directly treats the value function $p_w(\text{TP}|\mathbf{R})$ as a "reward" (Eq. (S6), see SI for more details). However, the performance of PGO strongly relies on the choice of the reward function, and is known to suffer issues including high variance and mode dropping (i.e., being trapped in sub-optimal solutions). In this regard, PGO could be less effective for complicated reactions (e.g., where multiple distinct transition states co-exist). Alternatively, to strike better balance between exploration and exploitation, we try to variationally optimize $V_\theta(\mathbf{R})$ w.r.t. a properly defined target distribution as in targeted adversarial learning optimized sampling (TALOS)[30] and variationally enhanced sampling (VES).[31] We term this approach as Variational Targeted Optimization (VTO), which is closely related to the soft Actor-Critic (SAC) algorithm[27] in RL. Specifically, since $-\log p_w(\text{TP}|\mathbf{R})$ can be interpreted as a kind of "energy", it naturally induces a distribution for exploitation, $p_{\text{exploit}} \propto p_w(\text{TP}|\mathbf{R})$. To encourage exploration, we also include an explorative distribution $p_{\text{explore}}$ (see SI for more details). The overall target distribution $p_T$ towards which $V_\theta(\mathbf{R})$ is optimized thus reads,

$$\log p_T = \alpha \log p_{\text{exploit}} + (1-\alpha) \log p_{\text{explore}} \quad (3)$$

where $0 \leq \alpha \leq 1$ is an inverse-temperature-like hyper-parameter trading-off exploration and exploitation: When $\alpha$ is close to zero, the sampling is dominant by exploration; While if $\alpha$ approaches unity, the sampling will focus merely on high-valued (TS-like) configurations. We can then minimize the Kullback-Leibler divergence (or relative entropy) between $p_\text{T}$ and $p_\text{shoot}$ (Eq. (S7); see SI for more details). In this way, the Expert is trained to mimic a "target policy" $p_\text{T}$. We tested both PGO and VTO in the following experiments. The workflow of RL$^\ddagger$ is illustrated in Fig. 1 and the assembled training protocol is summarized in SI and Algorithm S2.

## III. Results

In the following we will present evidence that RL$^{\ddagger}$ is able to automatically discover transition state(s) for reactions of various types. Specifically, we experimented RL$^{\ddagger}$ over several reactions with increasing complexity, including a textbook S$_N$2 reaction, a biomolecular conformation transition involvoving multiple transition states, as well as a Claisen rearrangement reaction complicated by solvent effects.

### 1. Numerical model potential

Before dealing with these complicated real-world reactions, we first illustrated how RL$^{\ddagger}$ works in an interpretable way on a model system, the Berezhkovskii-Szabo (BS) potential.[32] This 2-dimensional model potential consists of two local minima separated by an energy barrier (Fig. 2A), and captures key features of chemical reactions in a simplified manner. Two fully-connected ANNs are adopted to approximate the value function and the policy functions, respectively (see SI for more details about the model setup). We performed RL$^{\ddagger}$ for 100 EXIT's, and the policy function was optimized through PGO (see SI for training details). From Fig. 2B we can see that the training of RL$^{\ddagger}$ is very efficient, leading to a converged value function (with $\mathcal{L}(w)$ no longer diminishing significantly) within 50 EXIT's, and the average winning chance exceeds the prescribed threshold (10%) after 80 EXIT's. Besides, RL$^{\ddagger}$ is fairly sample-efficient in that only 10,000 random shootings were performed in total. After training is done, we plotted the policy function (Fig. 2C) and found it clearly diverts the sampling from the reactant or product regions to the TS region. This form of $V_\theta$ reminds one of the harmonic bias potential centered at the TS as adopted in umbrella sampling. However, unlike umbrella sampling, $V_\theta$ in RL$^{\ddagger}$ is learned without any knowledge of the reaction coordinate. On the other end, we can interpret the reaction mechanism based on the value function (Fig. 2D). As can be seen in Fig. 2D, the region of highest $p_w(\text{TP}|x,y)$ values corresponds exactly to the dividing surface of the BS potential. Additionally, we can track the training process of RL$^{\ddagger}$ by plotting the samples from $p_{\text{shoot}}$ at different training iterations (Fig. 2E). We find that the sampled configurations become increasingly concentrated around the TS region. Particularly after 100 EXIT's, the sampled configurations are located exactly around the TS, in agreement with the expected effect brought by the optimized policy $V_\theta$. This example demonstrates that RL$^{\ddagger}$ can be trained *tabula rasa* hence allows us to reveal the reaction mechanisms with minimal *a priori* expertise or assumptions.

### 2. S$_N$2 reaction

In the second experiment, RL$^{\ddagger}$ was applied to a textbook reaction, the substitution between Cl$^-$ and CH$_3$Cl (Fig. 3A), which is known to undergo a typical S$_N$2 mechanism. We can thus examine whether RL$^{\ddagger}$ is able to discover the transition state and unravel the correct mechanism of this chemical reaction from scratch. The two C-Cl bond distances were selected as order parameters $\mathbf{s} = (d_1, d_2)$ to define the reactant and product (Fig. 3A), and a reference free energy surface (FES) spanned over $\mathbf{s}$ is shown in Fig. 3B. This reaction was simulated by QM/MM MD in implicit solvent (see SI for simulation details). During EXIT training, we optimized a value function $p_w(\text{TP}|\mathbf{s})$ and policy function $V_\theta(\mathbf{s})$ which selectively input $\mathbf{s}$ rather than the coordinates of the whole system, and $V_\theta(\mathbf{s})$ is optimized via PGO (see SI for more details). RL$^{\ddagger}$ was trained to improve the ratio of successful shootings according to the final $p_{\text{shoot}}$ (Eq. (1)) exceeding 10% (see SI for more training details).

The optimized value function $p_w(\text{TP}|\mathbf{s})$ (Fig. 3C) sketches the TS regions in the lower-left diagonal regions of the $(d_1, d_2)$ space, indicating a concerted (i.e. S$_N$2) mechanism which agrees well with textbook knowledge. The bias potential $V_\theta(\mathbf{s})$ obtained by RL‡ (Fig. 3D) is again reminiscent of the harmonic potential commonly adopted in umbrella sampling, demonstrating that RL‡ is able to automatically create an external potential that effectively biases and concentrates the sampling over the TS regions. RL‡ reveals that the TS is quite symmetric in terms of the two C-Cl bonds (each is around 2.2 to 2.3 Å), and the -$CH_3$ motif takes a planar structure (Fig. 3E), as stated in the textbook. Furthermore, since sufficient samples from the TS ensemble were obtained by virtue of RL‡, we trained another value function $p_{w'}(TP|R)$ incorporating the positions of all atoms using SchNet[33] (see more details in SI). This value function allows us to investigate the reaction mechanism in atomistic details. Specifically, we defined the atom-wise "susceptibility" as the norm of the gradients of $p_{w'}(TP|R)$ w.r.t. the position of a certain atom (Eq. (S4); see SI for more details). According to this definition, atoms with relatively higher susceptibility can be considered to be more involved in the chemical transition (in other words, to be more relevant to the reaction coordinate). Consistent with the known mechanism, our susceptibility analysis (Fig. 3E) shows that the "reactive sites", namely, the two Cl atoms and the central carbon, on average display significantly higher susceptibility than the hydrogen atoms. Additonally, the TP ensemble yielded by RL‡ allows us to study the dynamics of the reaction. As an example, the transition path duration (TPD) provides valuable information about the kinetics and mechanisms of many chemical and bio-physical reactions[3, 34]. We collected the transition path ensemble of this chemical reaction (Fig. 3F), and analyzed the distribution of the TPD. Intriguingly, the distribution of TPD can be fitted by a log-normal distribution (Fig. S1) as in conformational transitions of some biomolecules recently detected through single-molecule experiments.[34]

## 3. Conformational transition of alanine dipeptide in explicit water

Next, we adopted RL‡ to study the conformational isomerization of alanine dipeptide (Ala2) in explicit water solvent, where the two backbone torsions, $\mathbf{s} = (\phi, \varphi)$, are slowly-changing variables governing the conformational changes (Fig. 4A). Since we are interested in the rotation of $\phi$, which is known to be slower than all the other backbone torsions,[22, 35] the reactant and the product are thus defined as the *cis-* and *trans-*conformations of $\phi$, respectively (see SI for more details about the definition). During EXIT, the value and policy functions are built upon $\mathbf{s}$. Unlike the previous examples, there are possibly more than one transition state connecting *cis-* and *trans-*$\phi$. In case of being locally trapped to a single TS, VTO was exploited to update the policy function during EXIT. Since we included a uniform distribution over $\mathbf{s}$ as the explorative component constituting the target distribution (Eq. 3; see SI for more simulation and training details), $V_\theta(\mathbf{s})$ enables efficient sampling over $\mathbf{s}$ in the early stages of EXIT. We thus performed density estimation according to the gathered samples (Fig. 4B). A total number of 240 EXIT's were performed, during which 24,000 random trajectories were shot, and the accumulated length of these trajectories was shorter than 5 ns. With this small computational overhead, the successful transition probability of the shooting moves according to the optimized $p_{shoot}$ (Eq. (1)) exceeds 11%.

After optimization is done, the value function $p_w(TP|\mathbf{s})$ was shown in Fig. 4C, and we can observe that $p_w(TP|\mathbf{s})$ peaks around where $\phi \approx 0°$, and the predicted TS region spreads over the entire range of $\varphi$, demonstrating the ability of RL‡ to unravel multiple transition states and transition pathways for complex reactions. The optimized policy function $V_\theta(\mathbf{s})$ (Fig. 4D) appears roughly complementary to the original FES of $(\phi, \varphi)$. This is resulted from a uniform distribution over $\mathbf{s}$ as the explorative component ($p_{explore}$) in the overall target distribution $p_T$. Upon closer inspection, we found that samples produced by $V_\theta$ centered around $\phi \approx 0°$, indicating that the final bias potential actually "creates" a deeper trench around the TS region, which is expected due to the exploitive part in $p_\text{T}$, i.e., $p_w(\text{TP}|\mathbf{s})$. Noteworthy, two different TS clusters with varying $\varphi$ values can be identified (Fig. 4E), yielding diverse transition paths initialized from these configurations.

Given the TS ensembles yielded by RL$^\ddagger$, we can re-examine the reaction mechanisms with finer details. As in the previous example, we trained a new value function $p_{w\prime}(\text{TP}|\mathbf{R})$ over the positions of all atoms (including Ala2 and the solvent). We also performed susceptibility analysis based on $p_{w\prime}(\text{TP}|\mathbf{R})$ (Fig. 4F). We found that atoms with the highest susceptibility in Ala2 center around the $C_\alpha$ atom, including the N-terminal peptide bond and the C-terminal carbonyl group. In fact, these atoms happen to be those defining the $\phi$ and $\varphi$ torsions, suggesting a concerted conformational transition mechanism between these two torsional angles. Remarkably, the atoms which can serve as hydrogen-bond acceptors or donators are of particularly high susceptibility (Fig. 4F), indicating the involvement of solvent coordinates in the transition pathway. $p_{w\prime}(\text{TP}|\mathbf{R})$ also allows us to investigate the solvation effects over the reaction. We performed susceptibility analysis over all water oxygen and hydrogen atoms, and calculated the radial averaged susceptibility (RAS; see Eq. (S12) for definition) for the solvent around Ala2. Figure 4G shows that the RAS rises around 2 to 3 Å and quickly decays to nearly zero within 4 Å, indicating that only water molecules within the first solvation shell may significantly contribute to the isomerization of $\phi$ torsion, possibly through the hydrogen bonds (as indicated by the earlier rise at <1.5 Å in the RAS of hydrogen atoms).

## 4. Claisen rearrangement in ionic liquid

In the last experiment, we implemented RL$^\ddagger$ to investigate a reversible Claisen rearrangement[36] in ionic liquid (Fig. 5A). This reaction involves a relatively high barrier thus brute-force simulations can hardly capture the chemical transitions. Textbooks usually categorize Claisen rearrangement as a typical concerted [3,3]-sigmatropic reaction. A plethora of theoretical studies of Claisen rearrangement in vacuum or implicit solvent have been recorded, most presuming a concerted mechanism and a 6-member-ring TS,[37] but less is known about how the solvent effects and thermal fluctuations would complicate the mechanism. The lengths of the two breaking/forming bonds were selected as order parameters $\mathbf{s} = (d_1, d_2)$ to define the reactant and product (Fig. 5A), and a value function $p_w(\text{TP}|\mathbf{s})$ along with a policy function $V_\theta(\mathbf{s})$ was optimized through EXIT. In the MD simulations, the reactant was treated by QM while the remaining solvent molecules (consisting of cations and anions, Fig. S2) were treated by MM. We adopted VTO for policy optimization, and included a uniform distribution over $\mathbf{s}$ as $p_\text{explore}$ in the overall target distribution (see SI for more details about the models, MD simulations and training procedures). Therefore, the FES over $\mathbf{s}$ can be estimated from the samples generated in the early stages of RL$^\ddagger$ as shown in Fig. 5B. We optimized the value and policy functions through EXIT till the successful shooting ratio surpassed 10% (see more training details in SI).

After training is done, we first inspected the value function. Different from what we found in the $S_N2$ reaction where the high value is limited to a relatively small range corresponding to concerted transitions (Fig. 2C), $p_w(\text{TP}|\mathbf{s})$ here (Fig. 5C) shows that a relatively high value spreads over a wide scope along the diagonal region of $\mathbf{s}$, indicating the possibility of non-concerted mechanisms. On the other hand, due to the explorative component in the target distribution, the bias potential $V_\theta(\mathbf{s})$ seems complementary to the FES (Fig. 5D). Intriguingly, when projecting those samples collected by the optimized $V_\theta$ to $\mathbf{s}$ as in Fig. 5B, we found that they dwell on at least two different regions. Since initial configurations from both regions have a relatively high chance of yielding a successful transition path according to the shooting results, we suppose that these two regions may correspond to two different transition states, one is a concerted TS (the blue rectangular in Fig. 5B) and the other is a stepwise TS (the yellow rectangular in Fig. 5B). Our findings suggest that the ionic liquid will complicate the reaction mechanism by allowing non-concerted reaction pathways. Indeed, water as a solvent was also reported to impose similar effects on this reaction.[14]

To further shed light on the reaction mechanism and the solvent effects, we trained a new value function $p_{w'}(\text{TP}|\mathbf{R})$ over the positions of all atoms (including the reactant and the solvent) based on the TS ensembles yielded by RL[‡] (see more details of $p_{w'}(\text{TP}|\mathbf{R})$ in SI). The susceptibility analysis over the reactant confirms that the bond breaking/forming sites constitute the "hotspots" for this reaction as expected (Fig. 5E). Besides, other heavy atoms in the reactant may also contribute to the chemical transitions to some extent (Fig. 5E). We found that the averaged TS is not a typical tight 6-member-ring structure in that the forming (or breaking) C-C bond ($d_1$) is significantly shorter than the breaking (or forming) C-O bond ($d_2$), suggesting a "late" TS for the forward reaction. Such a relatively loose TS would probably entail more significant charge separation,[14] and might result from the solvent effects. Motivated by this observation, we also computed the RAS of solvent molecules (Fig. 5F). Different from water as the solvent in the Ala2 example, in this case, the RAS of the ionic liquid does not decay until ~11 Å, suggesting that the ionic liquid imposes a relatively long-range effect over the reaction. This observation is consistent with that ionic liquid mainly interacts with the reactant through the Coulomb interaction, which is long-range in its physics nature. For example, we showcased a snapshot of the solvated TS structure in Fig. 5G, from which one can see that several solvent molecules form a "solvation cage" encompassing the reactant. Although not all of these molecular ions closely contact with the solute, they more or less interact with the reactant through their carried charges. How such a fluctuated electrostatic field may impact the reaction mechanism remains as an open question for further research.

## IV. Concluding remarks

In this paper, we presented a "shooting game" through molecular simulations in seek for the transition states of chemical reactions. A deep-learning-based approach was also introduced to play this game so as to unravel the underlying mechanisms. We named this machine-learning approach and the shooting game together as RL$^{\ddagger}$. RL$^{\ddagger}$ learns to propose hypothetic mechanisms, summarize feedbacks and gradually make self-corrections towards the optimal answers (that is, pinpoint the TS of a given reaction) in an end-to-end fashion. The feedback loop in RL$^{\ddagger}$ is formulated rigorously in the framework of reinforcement learning, so the virtual game player can directly learn from the previous experience (i.e., the simulation data). Therefore, RL$^{\ddagger}$ allows us to investigate reaction mechanisms in a more automatic manner and entail as least empirical presumptions as possible.

On the one hand, RL$^{\ddagger}$ benefits from knowledge-based methods: RL$^{\ddagger}$ is equipped with several state-of-the-art techniques developed in deep learning community, including Actor-Critic learning and EXIT, so its performance can be bootstrapped automatically meanwhile the training is very robust and efficient. On the other hand, RL$^{\ddagger}$ is physics-based in nature, given that the optimization objectives (i.e., the value and policy functions) of RL$^{\ddagger}$ are all formulated with clear physical interpretations: The value function measures the probability of a configuration being the TS, whereas the policy function is used to drive the molecular simulations more focused around the TS regions. In summary, RL$^{\ddagger}$ combines the strengths of both knowledge-based and physics-based methods, so it can efficiently simulate and extract mechanisms underlying chemical transitions as well as other rare events involving high (free) energy barriers.

**Acknowledgements**

The authors thank Wenjun Xie, Zacharias Faidon Brotzakis, Xing Che and Cheng-Wen Liu for useful discussion. This research was supported by National Natural Science Foundation of China [21927901, 21821004, 21873007 to Y.Q.G], the National Key Research and Development Program of China [2017YFA0204702 to Y.Q.G.] and Guangdong Basic and Applied Basic Research Foundation [2019A1515110278 to Y.I.Y.].

**Associated Content**

Supplemental Information consists of supplementary texts and figures. The derivation of RL$^{\ddagger}$ training objectives, algorithms, simulation setups and details are all wrapped in SI.

**Author contributions:** J.Z., Y.-K.L., Z.Z., L.Y. Y.I.Y and Y.Q.G. designed the research; J.Z., Y.-K.L. Z.Z., X.H., M.L. and Y.I.Y performed the research; J.Z., Y.-K.L., Z.Z., X.H., Y.I.Y and Y.Q.G. analyzed the data; J.Z., Y.-K.L., Z.Z., X.H., M.L., L.Y., Y.I.Y and Y.Q.G. wrote the paper.

**Notes:** The authors declare no conflict of interest.

**Figures**

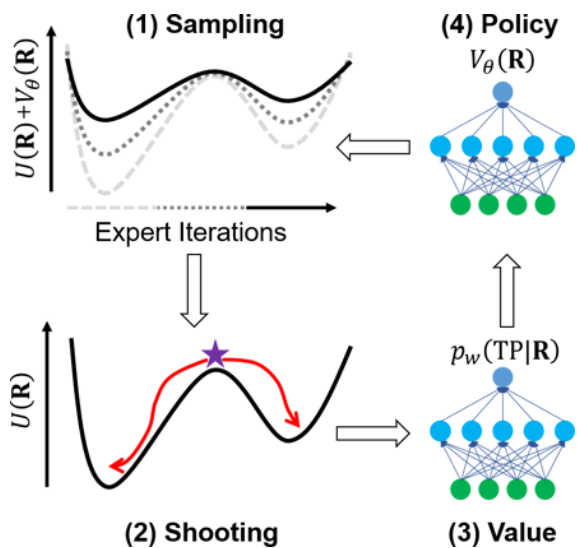

**Figure 1.** Workflow of RL$^\ddagger$. Each Expert Iteration in RL$^\ddagger$ consists of four steps: (1) Sampling on the modified PES (upper left), and different colors/symbols correspond to the PES at different iterations (the order of which is illustrated in the axis at the bottom); (2) Shooting over the original PES (lower left, where the purple star stands for the initial configuration from which two red-colored trajectories were launched); (3) Update the value function $p_w(\text{TP}|\mathbf{R})$; (4) Update the policy function (i.e., the bias potential) $V_\theta(\mathbf{R})$.

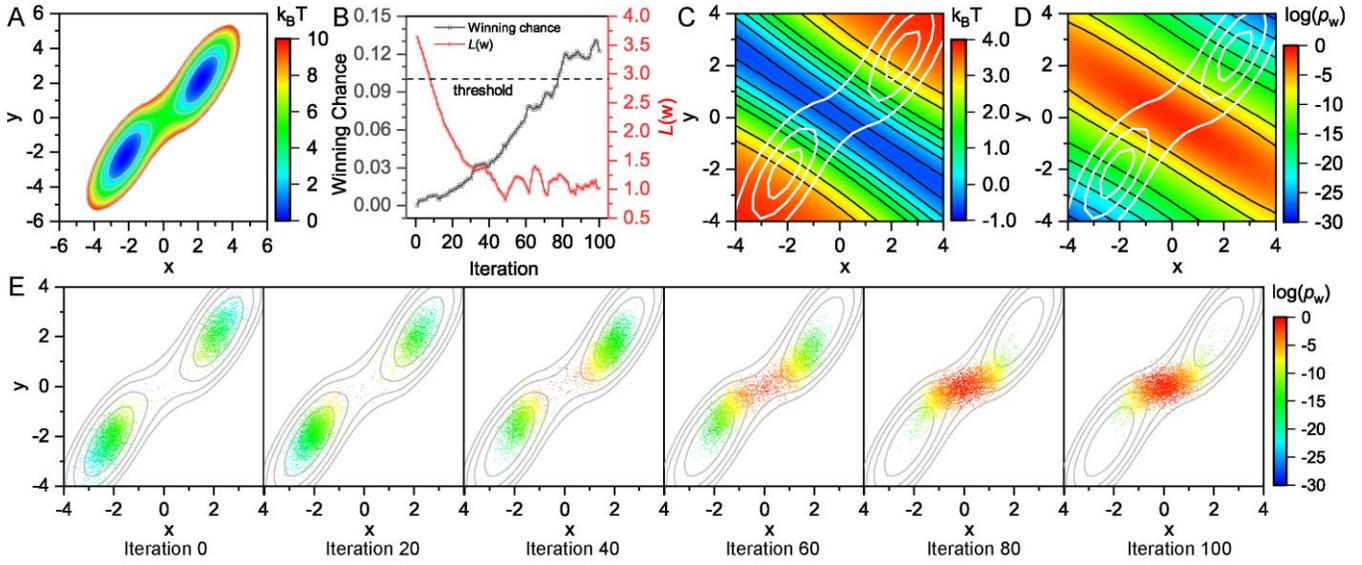

**Figure 2**. RL[‡] for BS model potential. (A) Colored contour map of the BS potential energy surface (PES). (B) Average winning chance (black line) and loss of the value function $\mathcal{L}(w)$ (red line) plotted against Expert Iterations. A typical stop-threshold for training (winning chance exceeds 10%) is shown in dashed line. Indeed, we trained the value and policy functions for 100 iterations. (C) Contour map of the final policy function $V_\theta(x, y)$. Transparent contour lines of the PES are shown in background. (D) Contour map of the final value function $\log p_w(\text{TP}|x, y)$. Transparent contour lines of the PES are shown in background. (E) Samples from $p_{\text{shoot}}$ at different Expert Iterations, colored according to the value function $\log p_w(x, y)$; Grey contour lines of the PES are shown in background.

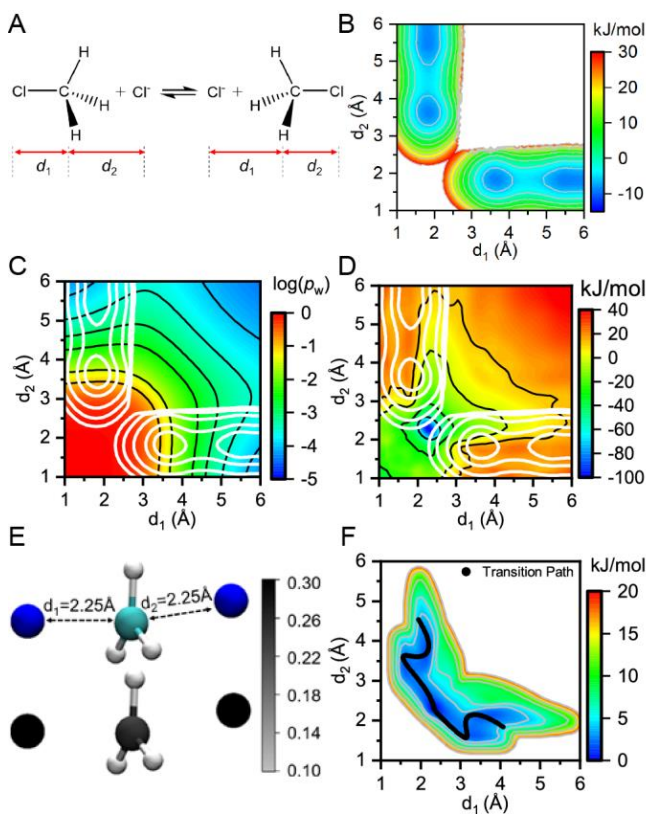

**Figure 3**. RL$^{\ddagger}$ for the S$_N$2 reaction. (A) Scheme of the reaction, where the two C-Cl bond distances are denoted by $d_1$ and $d_2$, respectively. (B) A contour plot for the reference free energy surface (FES) of the reaction. (C) A contour plot for the optimized value function $p_w(\text{TP}|\mathbf{s})$ in logarithm scale with $\mathbf{s} = (d_1, d_2)$. The reference FES of the reaction is shown as white contours in the background. (D) A contour plot for the optimized policy function $V_\theta(\mathbf{s})$. The reference FES of the reaction is shown as white contours in the background. (E) The upper panel shows a typical TS structure of the reaction (chlorides are rendered in blue, carbon in cyan and hydrogens in white), and the average values of $d_1$ and $d_2$ for TS are numbered. In the lower panel, the same structure is colored according to the averaged susceptibility of $p_{w'}(\text{TP}|\mathbf{R})$, with the color bar shown on the right. (F) The heat map of the transition path ensemble projected on $(d_1, d_2)$, superimposed with a typical transition path spanned by solid black dots.

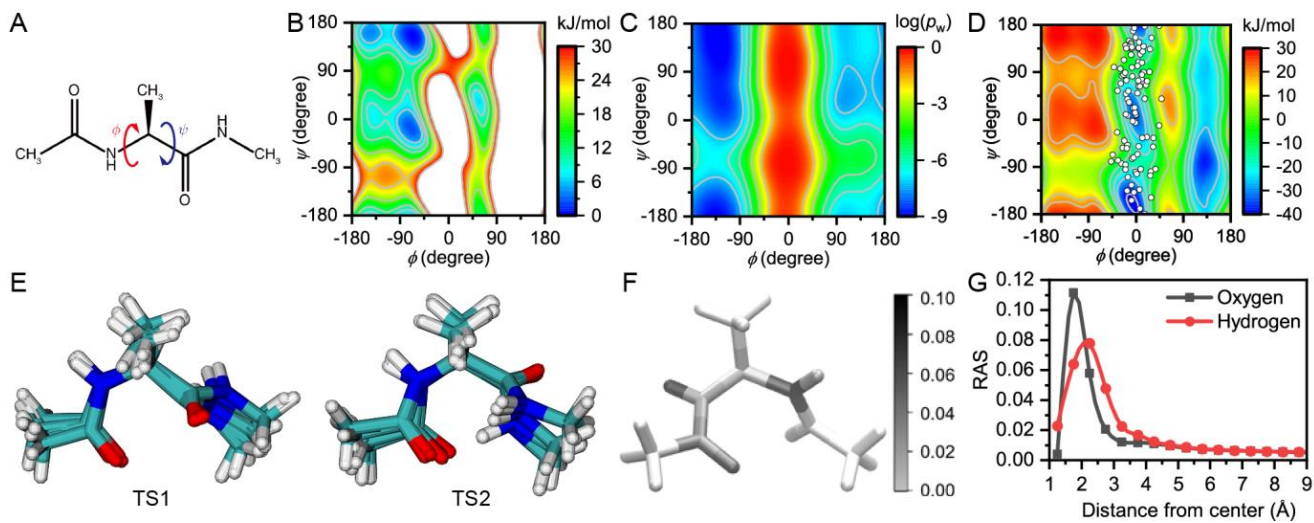

**Figure 4.** RL$^{\ddagger}$ for Ala2 in explicit water. (A) Chemical structure of Ala2 and two collective variables: the torsions $\phi$ and $\varphi$. (B) A reference FES for Ala2 over collective variables $\mathbf{s} = (\phi, \varphi)$. (C) Contour map of the optimized value function shown in logarithm, $\log p_w(\text{TP}|\mathbf{s})$. The maximal value is shifted to zero during post-processing. (D) Contour map of the optimized policy function $V_\theta(\mathbf{s})$, superimposed with samples (white circled dots) under simulations biased by $V_\theta(\mathbf{s})$. (E) Structures of two typical transition states. TS1 adopts $(\phi \approx 0°, \varphi \approx 90°)$ whereas TS2 adopts $(\phi \approx 0°, \varphi \approx -90°)$. (F) The average susceptibility of $p_w(\text{TP}|\mathbf{R})$ for the solute atoms. The presented molecule is similarly posed as the TS2 in (E). (G) The radial averaged susceptibility (RAS) of $p_{w\prime}(\text{TP}|\mathbf{R})$ for the solvent atoms (oxygen atoms is shown in black line and symbols whereas hydrogen atoms in red).

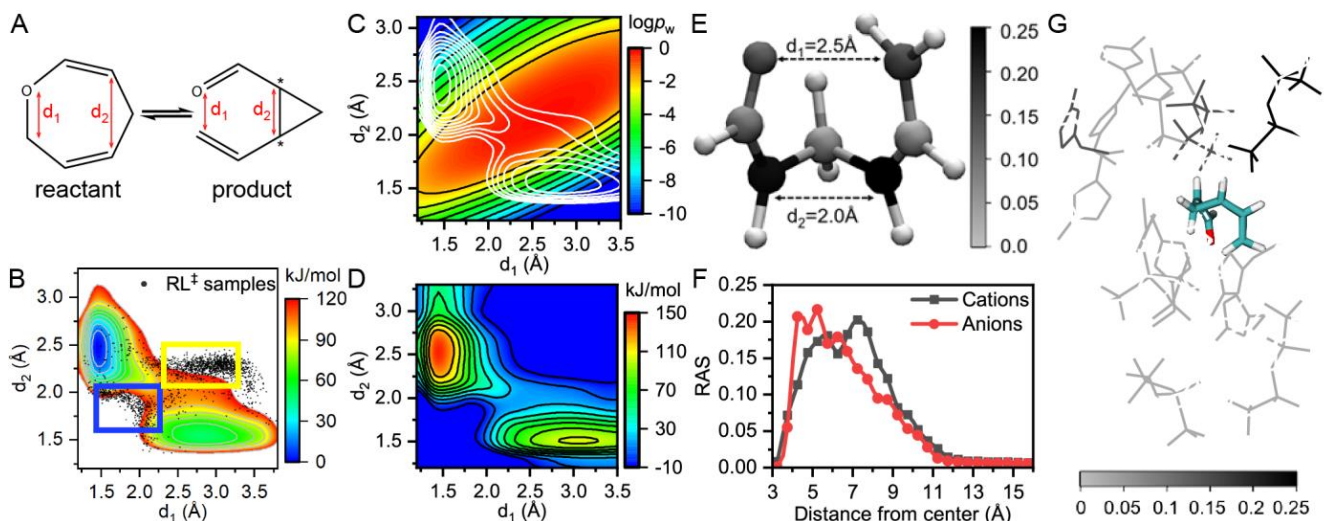

**Figure 5.** RL$^{\ddagger}$ for a reversible Claisen rearrangement in ionic liquid. (A) Scheme of the reaction under study. The lengths of two breaking/forming bonds are denoted as $d_1$ and $d_2$, respectively. The forming C-C bond leads to two chiral carbons which are denoted by asterisks. (B) The FES of the reaction (estimated according to the samples obtained during RL$^{\ddagger}$), superimposed with representative samples (black dots) generated by the optimized policy function which spread over two different regions: the concerted TS region (within the blue rectangular) and the stepwise TS region (within the yellow rectangular). (C) A contour plot for the optimized value function $p_w(\text{TP}|\mathbf{s})$ in logarithm scale with $\mathbf{s} = (d_1, d_2)$. The reference FES is shown as white contours in the background. (D) A contour plot for the optimized policy function $V_\theta(\mathbf{s})$. (E) A typical TS structure is shown with the average values of $d_1$ and $d_2$ marked out. Each atom is colored according to the susceptibility of $p_{w'}(\text{TP}|\mathbf{R})$ averaged over the TS ensemble. (F) The radial averaged susceptibility (RAS) for the solvent molecules (cations are shown in black whereas anions in red). (G) A snapshot of solvated reactant. The reactant is colored according to elements (carbons in cyan, oxygen in red and hydrogens in white). The solvent molecules are colored according to the instantaneous susceptibility $p_{w'}(\text{TP}|\mathbf{R})$ with the color bar shown in the bottom. Note that many molecules are in white (i.e., with zero susceptibility) thus invisible.